\documentclass[conference]{IEEEtran}
\IEEEoverridecommandlockouts
\usepackage{cite}
\usepackage{amsmath,amssymb,amsfonts}
\usepackage{algorithmic}
\usepackage{graphicx}
\usepackage{textcomp}
\usepackage{xcolor}
\usepackage{subcaption}
\def\BibTeX{{\rm B\kern-.05em{\sc i\kern-.025em b}\kern-.08em
    T\kern-.1667em\lower.7ex\hbox{E}\kern-.125emX}}
\begin{document}

\title{The Rich Get Richer in Bitcoin Mining\\Induced by Blockchain Forks
}

\author{\IEEEauthorblockN{Akira Sakurai}
\IEEEauthorblockA{\textit{Kyoto University} \\
Kyoto, Japan }
\and
\IEEEauthorblockN{Kazuyuki Shudo}
\IEEEauthorblockA{\textit{Kyoto University} \\
Kyoto, Japan }
}

\maketitle

\begin{abstract}
Bitcoin is a representative decentralized currency system. For the security of Bitcoin, fairness in the distribution of mining rewards plays a crucial role in preventing the concentration of computational power in a few miners. Here, fairness refers to the distribution of block rewards in proportion to contributed computational resources. If miners with greater computational resources receive disproportionately higher rewards—i.e., if the ``Rich Get Richer'' (TRGR) phenomenon holds in Bitcoin—it indicates a threat to the system’s decentralization.
This study analyzes TRGR in Bitcoin by focusing on unintentional blockchain forks, an inherent phenomenon in Bitcoin. Previous research has failed to provide generalizable insights due to the low precision of their analytical methods. In contrast, we avoid this problem by adopting a method whose analytical precision has been empirically validated.
The primary contribution of this work is a theoretical analysis that clearly demonstrates TRGR in Bitcoin under the assumption of fixed block propagation delays between different miners. More specifically, we show that the mining profit rate depends linearly on the proportion of hashrate. Furthermore, we examine the robustness of this result from multiple perspectives in scenarios where block propagation delays between different miners are not necessarily fixed.
\end{abstract}

\section{Intruduction}
Bitcoin\cite{bitcoin} is a peer-to-peer currency system that enables transactions without relying on a trusted third party. In the Bitcoin protocol, nodes that process transactions are referred to as miners. Anyone can become a miner, and miners are incentivized to follow the protocol, thereby eliminating the need for users to trust any centralized authority when processing transactions.

From a security perspective, it is essential that computational resources (i.e., hashrate) are not concentrated in Bitcoin. Concretely, the protocol assumes that the majority of computational resources are held by honest miners in order for transactions to be securely processed. For example, if a particular miner gains control of more than 51\% of the network's total hashrate, they could arbitrarily alter transaction histories\cite{BitcoinBackboneProtocol}\cite{EverythingisaRace}\cite{TightConsistencyBoundsforBitcoin}. Even when this threshold is not surpassed, attacks such as selfish mining\cite{majorityisnotenough} become increasingly viable as a miner's proportion of the hashrate grows. Conversely, under certain conditions, it has been shown that following the Bitcoin protocol is economically rational when the hashrate distribution is sufficiently decentralized\cite{BlockchainMiningGames}.

To suppress excessive concentration of hashrate, it is crucial for mining to be fair. Here, we define mining fairness as the condition in which miners receive block rewards in proportion to the computational resources they commit. Unfairness in mining implies that certain miners receive greater rewards than others despite contributing proportionally similar computational effort. In particular, if miners with larger hashrates receive disproportionately higher block rewards—a phenomenon known as the Rich Get Richer (TRGR)—it suggests that the Bitcoin protocol inherently promotes hashrate centralization. For example, miners are incentivized to shift their hashrate from smaller mining pools to larger ones. This highlights a fundamental design flaw in Bitcoin.

In this study, we analyze the structure of TRGR in Bitcoin by investigating the impact of unintentional blockchain forks on mining fairness. Forks are an essential and unavoidable phenomenon in Bitcoin. Bitcoin was originally proposed as a highly decentralized system, in which miners act as independent agents and are geographically dispersed. As a result, Bitcoin functions as a distributed system, and its blockchain must be synchronized across the network whenever new blocks are generated. A blockchain fork occurs when a new block is generated before the previously mined block has been fully propagated throughout the network, making it an inherent part of the protocol’s operation.

Although some prior work\cite{ImpactofTemporaryFork} has suggested the existence of TRGR, their analyses have relied on methods with insufficient accuracy. Specifically, the estimated impact of forks on mining fairness has been shown to deviate from actual values by more than 100\%\cite{sakurai2025modelbasedanalysisminingfairness}. In this paper, we overcome this limitation by employing the model-based method proposed by Sakurai et al.\cite{sakurai2025modelbasedanalysisminingfairness}, which significantly improves analytical accuracy by formally treating forks using time intervals called "rounds." This approach enables a precise assessment of the impact of forks on mining fairness.

Our main contributions in this study are as follows:
\begin{itemize}
  \item \emph{We theoretically demonstrate the existence of the Rich Get Richer (TRGR) phenomenon in Bitcoin under the assumption of fixed block propagation delays between different miners.} (Section \ref{fixed analysis})
  \begin{itemize}
    \item We confirm that the mining profit rate increases monotonically and linearly with a miner’s proportion of the total hashrate.
    \item We mathematically establish the trade-off between decentralization and transaction processing capacity through mining fairness. This result implies that improving block propagation delays can enhance mining fairness.
    \item We demonstrate that the break-even point for mining profit rate increases as the hashrate distribution becomes more centralized. Specifically, we confirm that the break-even threshold is equal to the sum of the squares of the miners' hashrate proportions. This quantity increases as the distribution becomes more imbalanced. This implies that profitability becomes harder to achieve for all miners as the hashrate becomes more concentrated.
    \item We show that the influence of the tie-breaking rule is relatively small compared to the effects of block propagation delay and block generation interval.
  \end{itemize}

  \item We further validated and analyzed TRGR in Bitcoin when block propagation delays are not necessarily fixed. (Section \ref{variable analysis})
  \begin{itemize}
    \item By comparing the fixed-delay scenario with one where block propagation times are randomly distributed between different miners, we show that TRGR still holds on average. Furthermore, we demonstrate that the effect of propagation delay is more pronounced for small-scale miners with low hashrate.
    \item Under a simplified setting in which miners can strategically adjust their block propagation delays, we showed that TRGR persists at Nash equilibrium.
  \end{itemize}
\end{itemize}

\section{Background}
\subsection{Bitcoin}
In Bitcoin, users initiate transactions when they use the currency. These transactions are broadcast over the Bitcoin peer-to-peer (P2P) network and are grouped into blocks by nodes known as miners.

Each block references a single parent block, and this referencing structure forms a chain of blocks, known as the blockchain. Each miner maintains its own local view of blocks, and the longest chain is regarded as the valid transaction history, commonly referred to as the main chain.

To generate a block, a miner constructs a block template containing information such as the processed transactions, a timestamp, and the hash of the latest block on the main chain. The miner then performs repeated hash computations until the hash of the block header falls below a specified threshold. This process is known as mining. A block that satisfies this condition is considered valid and is propagated across the network. Upon verifying the block’s validity, each miner updates its own main chain. Only blocks included in the main chain are considered valid transaction results, and the miner who generates such a block receives a block reward. This reward consists of a base reward and transaction fees. 

An attacker invalidates a processed transaction by mining blocks that override the block containing the target transaction. Specifically, the attacker initiates a fork and constructs a new blockchain that is longer than the one in which the target transaction was included. To successfully carry out such an attack with certainty, the attacker would require computational power equivalent to the total network hashrate, which is generally considered prohibitively expensive and unrealistic. In this sense, the security of Bitcoin is ensured.

Even when all miners follow the Bitcoin protocol honestly, it is still possible for the blockchain to fork during block propagation. A fork occurs when a new block is generated by a miner before a previously mined block has been fully propagated throughout the network. The likelihood of a fork increases with block propagation delay\cite{informationpropagation}. In this study, we focus exclusively on such unintentional forks and do not consider intentional ones.

Each miner selects the longest blockchain as its main chain for mining. If there is ambiguity in identifying the longest chain due to a fork, a tie-breaking rule is used. In Bitcoin, the miner adopts the chain it received first—a policy known as the first-seen rule. This rule is used in practice due to its simplicity and effectiveness. However, from the perspective of selfish mining, it has a drawback: the overall security of the system becomes highly dependent on the attacker’s block propagation capability\cite{majorityisnotenough}. To address this issue, alternative tie-breaking rules have been proposed, such as the random rule\cite{majorityisnotenough}, which selects a chain at random, and the last-generated rule\cite{sakurai2024fullylocallastgeneratedrule}\cite{sakurai2024tiebreaking}\cite{oneweirdtricktostopselfishminers}, which selects the chain containing the most recently generated block.

\subsection{Mining Fairness}
We define mining fairness as a state in which each miner receives block rewards in proportion to the computational resources (hashrate) they contribute to mining. Since blocks are generated in proportion to hashrate and the Bitcoin protocol adjusts mining difficulty to ensure that the total computational resources in the network match the rate of block generation, mining fairness holds as long as no forks occur. However, as discussed above, forks occur probabilistically, meaning not all blocks are included in the main chain. Consequently, some blocks receive no reward, and mining fairness is violated.

The impact of blockchain forks on mining fairness is influenced by several factors. The probability that a fork occurs depends on the block’s propagation delay and the overall distribution of hashrate. For example, forks are less likely when propagation delays are short and the hashrate distribution is biased. Furthermore, whether a block involved in a fork is ultimately included in the main chain depends on factors such as the tie-breaking rule, propagation delay, and the block generator’s proportion of the hashrate.

Understanding how forks affect mining fairness is not straightforward. For instance, a miner with a larger proportion of the hashrate is less likely to cause a fork when generating a block. However, this does not necessarily mean the miner will receive a larger reward. In selfish mining, the root cause of excessive block rewards for the attacker is the ability to invalidate blocks generated by honest miners. A miner with a large hashrate proportion has fewer opportunities to invalidate others’ blocks since their own blocks are less likely to result in forks. On the other hand, when a fork does occur, blocks generated by miners with a large hashrate proportion are more likely to be included in the main chain. This increases their expected block rewards and contributes to the TRGR effect.

The aim of this study is to analyze the impact of unintentional blockchain forks on mining fairness and to determine whether TRGR holds in Bitcoin. To ensure decentralization, Bitcoin relies on independent miners that are geographically distributed. In such a system, blockchain forks are unavoidable due to block propagation delays.

While other factors may also affect mining fairness—such as ASIC performance or electricity costs—these are not considered in this study, as their influence is comparatively straightforward to understand.

\section{Related Work}
Chen et al.\ examined the impact of unintentional blockchain forks on mining fairness in Bitcoin\cite{ImpactofTemporaryFork}. One of their conclusions is that TRGR phenomenon holds in Bitcoin. However, their analytical method suffers from low precision and limited generality. Specifically, they did not consider the effect of forks on the rate at which miners initiate new rounds, nor the increase in block rewards due to fork creation. According to the study by Sakurai et al.\cite{sakurai2025modelbasedanalysisminingfairness}, neglecting the effect of forks on the proportion of round initiation alone can result in over 100\% error in estimating actual mining fairness.

Attacks that intentionally exploit forks—such as selfish mining\cite{majorityisnotenough}, fork-after-withholding\cite{FAW}—also suggest the existence of TRGR in Bitcoin. For instance, in selfish mining, the attacker's block reward increases with the attacker’s hashrate proportion. Further studies that incorporate the effect of stale blocks into selfish mining\cite{onTheSecurity, StochasticModellingofSelfishMininginProof-of-WorkProtocols} suggest that longer block propagation times intensify the TRGR effect.

We now turn to studies that investigate TRGR in terms of hashrate distribution, without considering forks. Judmayer et al.\ showed that a small number of mining pools control the majority of the network’s hashrate by attributing each block to its generator pool\cite{MergedMining}. Romiti et al.\ refined this attribution method to improve accuracy and demonstrated, using the Gini coefficient, that the hashrate distribution is highly concentrated\cite{deepdiveintoBitcoin}. Cong et al.\ argued that large mining pools do not necessarily grow further, since individual miners can split their hashrate across multiple pools and large pools tend to impose higher fees\cite{DecentralizedMininginCentralizedPools}.
Huang et al.\ investigated whether TRGR arises in blockchain systems\cite{DotheRichGetRicher}. They primarily showed that TRGR appears in Proof-of-Stake-based systems, but concluded that it does not occur in Proof-of-Work systems. This conclusion, however, stems from their failure to account for the impact of blockchain forks.
Li et al.\ argued, using a mean-field game model, that reward instability itself can give rise to TRGR\cite{AMeanFieldGamesModel}.

Next, we review studies that analyze TRGR from the perspective of wealth distribution. Ron et al.\ examined transaction data on the blockchain and found that wealth distribution in Bitcoin is extremely skewed\cite{QuantitativeAnalysisoftheFullBitcoin}. They also identified that a significant proportion of Bitcoin's transaction volume originates from a single transaction issued in 2010. From a network science perspective, Kondor et al.\ demonstrated signs of TRGR by analyzing transaction histories\cite{KondorDotheRichGetRicher, KondorDotheRichStillGetRicher}. Similar trends were also confirmed by Gupta et al.\cite{GiniGupta}, Maesa et al.\cite{Data-drivenanalysis}, and Venturini et al.\cite{vmappingnetworkstructuresdynamics}.
Sai et al.\ examined multiple cryptocurrency systems and found that while systems with large market capitalizations tend to be relatively more decentralized, wealth concentration remains\cite{CharacterizingWealthInequality}. Juodis et al.\ extended the analysis to Layer-2 cryptocurrency systems and introduced the Herfindahl–Hirschman Index (HHI) as a metric\cite{Overviewandempiricalanalysisofwealthdecentralization}. Kusmiers et al.\ compared ERC-20 tokens\cite{eip20} with Bitcoin and showed that ERC-20 tokens tend to exhibit higher centralization\cite{Howcentralizedisdecentralized}.

\section{Model}
We describe the Bitcoin network model used in this study, following Sakurai et al.\cite{sakurai2025modelbasedanalysisminingfairness}. We assume that all miners are honest and follow the Bitcoin protocol. The set of miners is denoted by $V$, which is fixed throughout the analysis. Each miner $i \in V$ is assigned a hashrate proportion $\alpha_i$, satisfying $\sum_{i \in V} \alpha_i = 1$. We assume that at most two blocks can be generated per round. A round is defined as a global time interval, specifically, the period between the generation of the first block at height $r$ and the generation of the first block at height $r+1$. This round-based model of the blockchain network enables us to formally handle forks and precisely capture their impact on mining fairness.

A fork is defined as the event in which two blocks are generated within a single round. Let $F_{ij}$ denote the probability that a fork occurs when miner $i$ initiates a round and subsequently miner $j$ generates a block. Let $W_{ij}$ denote the probability that, in such a fork, the block generated by miner $i$ is included in the main chain. We assume that the block reward is a fixed value across all blocks.

\section{Method for Calculating Mining Profit Rate}
In this study, we use the mining profit rate as an indicator to analyze mining fairness.
Mining profit rate refers to the mining profit earned per unit of computational resource. In this section, we explain the method proposed by Sakurai et al.\cite{sakurai2025modelbasedanalysisminingfairness} for calculating mining profit rate. Their method performs this calculation within a round-based blockchain network model, providing a high-speed and accurate alternative to simulation-based approaches. Simulation results have verified that this method significantly improves the accuracy of mining fairness analysis compared to existing techniques.

The parameters required to compute the mining profit rate are: the hashrate proportion $\alpha_i$ of each miner $i$, the block propagation delay $T_{ij}$ from miner $i$ to miner $j$, and the average block generation interval $T$. The propagation delay $T_{ij}$ is defined as the time it takes for a block generated by miner $i$ to reach miner $j$.

We first calculate the fork rate $F_{ij}$, which is the probability that miner $j$ causes a fork in the same round under the condition that miner $i$ initiates the round and miner $j$ is the next to generate a block. In blockchain networks, block generation intervals follow an exponential distribution. Therefore, the probability that the next block is generated before the previous one reaches other miners corresponds to the fork probability, and is given by:
\begin{align}
  F_{ij} = 1 - \exp\left(-\frac{T_{ij}}{T}\right)
\end{align}

Next, we compute the probability $W_{ij}$ that miner $i$’s block is included in the main chain when a fork occurs between miners $i$ and $j$. This probability can be approximately calculated based on the tie-breaking rule, the hashrate distribution, and the propagation delay. For details of the approximation method, we refer the reader to the study by Sakurai et al.\cite{sakurai2025modelbasedanalysisminingfairness}.

We then compute the proportion of round initiation for each miner. The proportion of round initiation of a miner is the probability that the miner generates the block that starts a new round—equivalently, the probability that the blockchain height is updated by that miner. Let $X_r$ be the random variable representing the miner who initiates round $r$. Then, the following recurrence holds:
\begin{align}
  P(X_{r+1} = i) = \sum_{j \in V} \left(\alpha_i (1 - F_{ji}) + \sum_{k \in V} \alpha_k F_{jk} \alpha_i \right) P(X_{r} = j) \label{rounditerate}
\end{align}
Since this process forms an ergodic Markov chain, its limit distribution converges to a stationary distribution. By iterating the above equation, we can obtain the proportion of round initiation $\pi_i$ for each miner $i$.

Once $\pi_i$ is computed, the proportion of block reward $r_i$ for miner $i$ is given by:
\begin{align}
r_i =& \pi_i\left(1 - \sum_{j \in V} \alpha_j F_{ij} + \sum_{j \in V} \alpha_j F_{ij} W_{ij} \right) \nonumber\\
    &+ \sum_{j \in V} \pi_j \alpha_i F_{ji} (1 - W_{ji}) \label{blockreward}
\end{align}

From the proportion of block reward, we can compute the mining profit $MP_i$ and mining profit rate $MPR_i$ for miner $i$ as follows:
\begin{align}
  MP_i &= r_i - \alpha_i \\
  MPR_i &= \frac{MP_i}{\alpha_i}
\end{align}
In Bitcoin, the block generation difficulty is adjusted approximately every two weeks. This adjustment ensures that the total computational effort spent on mining matches the total rewards distributed. As a result, by comparing a miner’s hashrate proportion $\alpha_i$ with its proportion of block reward $r_i$, we can evaluate the mining profit and mining profit rate of that miner.

\section{Fixed Block Propagation Delay Between Different Miners} \label{fixed analysis}
We analyze the structure of the TRGR effect in Bitcoin under the condition that block propagation delays between different miners are fixed. First, in Section~\ref{theoretical analysis}, we derive an approximate theoretical formula for mining profit rate. Next, in Section~\ref{verification}, we verify this formula through numerical computation. Finally, we discuss the insights obtained from the derived approximation in Section~\ref{insight}.

\subsection{Theoretical Analysis} \label{theoretical analysis}
We perform a theoretical analysis that does not rely on specific parameter values.

The block propagation delay between any two distinct miners is fixed at $d$. We assume that the ratio $d/T$ between the propagation delay $d$ and the average block generation interval $T$ is sufficiently small. Under these conditions, the fork rate $F_{ij}$ is given by:
\begin{align}
  F_{ij} = 
  \begin{cases}
    0 & \text{if } i = j, \\
    1 - \exp(-\frac{d}{T}) & \text{if } i \neq j.
  \end{cases}
\end{align}
Hereafter, we denote $F_{ij}$ by $f$ for $i \neq j$. Since $f \approx d/T$, we can treat $f$ as a sufficiently small parameter.

We first compute the proportion of round initiation $\pi_i$ for miner $i$. From Equation~\ref{rounditerate}, we obtain:
\begin{align}
  \pi_i &= \sum_{j \in V} \left(\alpha_i (1 - F_{ji}) + \sum_{k \in V} \alpha_k F_{jk} \alpha_i \right) \pi_j \\
        &= \alpha_i + \alpha_i f (\alpha_i - \sum_{j \in V} \alpha_j \pi_j)
\end{align}
By substituting $\pi_j = \alpha_j$ on the right-hand side for all $j$, we get:
\begin{align}
  \pi_i \approx \alpha_i + \alpha_i f (\pi_i - \sum_{j \in V} \alpha_j^2) \label{roundstartrateapprox}
\end{align}
This gives an approximate expression for the proportion of round initiation of each miner $i$. Notably, this result is independent of the tie-breaking rule.

Recalling that the block propagation delay between different miners is fixed, the value of $W_{ij}$ depends on the tie-breaking rule and is given by:
\begin{align}
  W_{ij} = 
  \begin{cases}
    1 - \alpha_j & \text{first-seen rule,} \\
    \frac{1 - \alpha_j + \alpha_i}{2} & \text{random rule,} \\
    \alpha_i & \text{last-generated rule.}
  \end{cases}
\end{align}
For instance, under the first-seen rule, the block that initiates the round reaches all miners other than miner $j$ before the forked block, so these miners mine on the block generated by miner $i$.

Next, we calculate the proportion of block reward $r_i$ for miner $i$. From Equation~\ref{blockreward}, we obtain:
\begin{align}
r_i &= \pi_i\left(1 - \sum_{j \in V} \alpha_j F_{ij} + \sum_{j \in V} \alpha_j F_{ij} W_{ij}\right) \nonumber\\
    &\quad + \sum_{j \in V} \pi_j \alpha_i F_{ji} (1 - W_{ji}) \\
    &= \alpha_i + 2 \alpha_i f (\alpha_i - \sum_{j \in V} \alpha_j^2) + \mathcal{O}(f^2) \\
    &\approx \alpha_i + 2 \alpha_i f (\alpha_i - \sum_{j \in V} \alpha_j^2)
\end{align}
Therefore,
\begin{align}
r_i &= \alpha_i + 2 \alpha_i f (\alpha_i - \sum_{j \in V} \alpha_j^2) \\
\iff MPR_i &= \frac{r_i - \alpha_i}{\alpha_i} = 2 f (\alpha_i - \sum_{j \in V} \alpha_j^2) \label{mprapprox}
\end{align}
The approximation in Equation~\ref{mprapprox} is justified by the assumption that $f$ is sufficiently small. Moreover, this result does not depend on the tie-breaking rule.

\begin{figure}[tb]
  \centering
  \includegraphics[width=\linewidth]{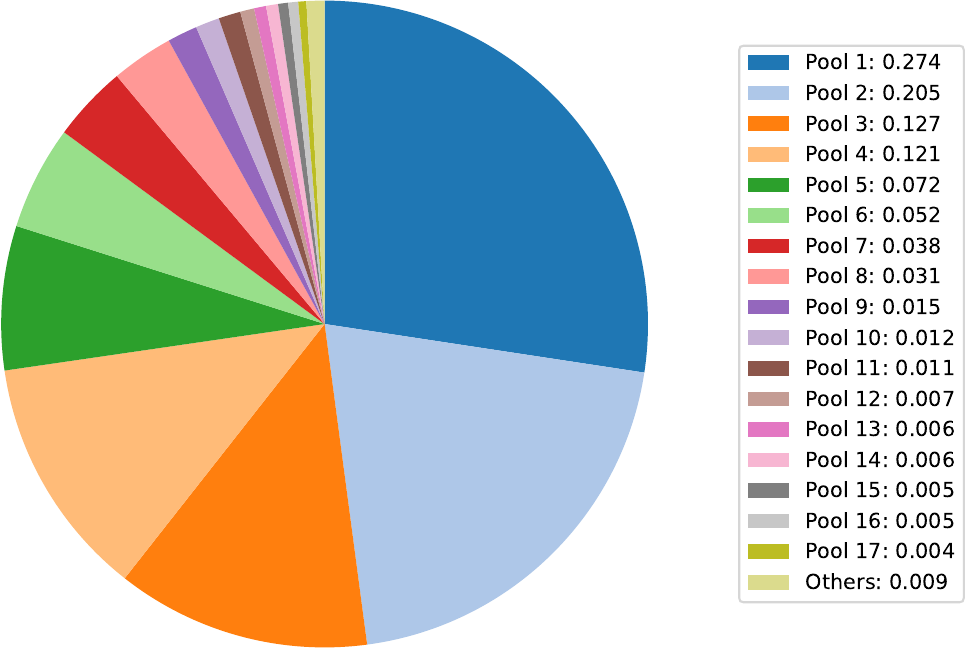}
  \caption{Hashrate distribution in Bitcoin.}
  \label{bitcoindis}
\end{figure}

\begin{figure}[tb]
  \centering    
  \includegraphics[width=\linewidth]{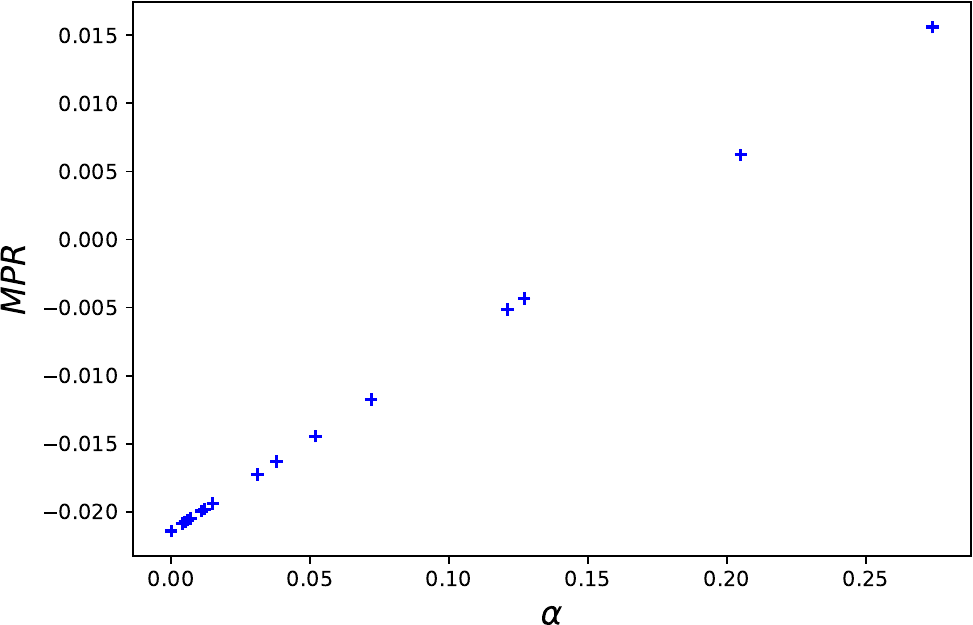}
  \caption{Numerical results for mining profit rate.}
  \label{numresult}
\end{figure}

\subsection{Validation via Numerical Computation} \label{verification}
In the preceding analysis, two approximations were employed, specifically those in Equations~\ref{roundstartrateapprox} and~\ref{mprapprox}. Here, we verify the validity of these approximations. Concretely, we substitute actual parameter values used in the method by Sakurai et al.~\cite{MiningPoolStats} and compare the numerical results with those obtained from our theoretical analysis. If the numerical results agree with the theoretical ones, and considering that these numerical results closely reflect the actual values observed in Bitcoin, we can conclude that our theoretical analysis is applicable to Bitcoin.

The specific parameters used were as follows: the number of miners was set to $1000$, the average block generation interval was $T = 600$, and the hashrate distribution was based on the actual distribution observed in Bitcoin~\cite{MiningPoolStats} (see Figure~\ref{bitcoindis}). To be precise, hashrates were assigned to miners to match the distribution shown in the figure, and the remaining hash power was evenly distributed among the rest of the miners. We considered three tie-breaking rules: the first-seen rule, the random rule, and the last-generated rule.

For the block propagation delay $d$ between different miners, we considered values up to $42$. This choice was motivated by three reasons. First, from the nature of the approximation, the smaller the block propagation delay between different miners, the more accurate the approximation becomes. Thus, our theoretical result remains applicable for $d \leq 42$. Second, the ratio $d/T$ is lower than those found in most blockchain systems, including Ethereum, which has a small block interval (approximately $d/T = 0.7$~\cite{onTheSecurity})~\cite{Calibratingtheperformance}. Lastly, Sakurai et al.'s method has been validated using simulations for values of $d/T$ up to $0.1$.

Figure~\ref{numresult} presents the numerical results for the mining profit rate $MPR_i$ under the first-seen rule. As in Equation~\ref{mprapprox}, the results demonstrate that $MPR_i$ depends linearly on the miner’s proportion of the total hashrate. The average correlation coefficient across the three tie-breaking rules is $0.999992$, indicating that this linear dependency holds irrespective of the choice of tie-breaking rule.

A line is determined by its slope and a point on the line. We first compare the slopes. Table~\ref{comparisonforksloop} presents the theoretically derived slope $2f$ alongside the results from numerical computation under the first-seen rule. The theoretical slope closely matches the numerical one.

Next, we compare the zero-point of the mining profit rate. Since $MPR_i$ is linearly dependent on $\alpha_i$, the following holds:
\begin{align}
MPR_i &= k (\alpha_i - \alpha_0), \label{basef}
\end{align}
where $k$ is the slope of the line, $\alpha_i$ denotes the hashrate proportion of miner $i$, and $\alpha_0$ denotes the zero-point. We now demonstrate that this zero-point equals $\sum \alpha_i^2$. From Equation~\ref{basef}, we obtain:
\begin{align}
MPR_i \alpha_i &= \alpha_i k (\alpha_i - \alpha_0), \\
\iff MP_i &= k \alpha_i^2 - k \alpha_0 \alpha_i.
\end{align}
Summing over all miners yields:
\begin{align}
\sum_{i \in V} MPR_i &= \sum_{i \in V} \left( k \alpha_i^2 - k \alpha_0 \alpha_i \right) \\
\iff 0 &= k \sum_{i \in V} \alpha_i^2 - k \alpha_0 \sum_{i \in V} \alpha_i \\
\iff \alpha_0 &= \sum_{i \in V} \alpha_i^2.
\end{align}
Thus, the zero-point of the mining profit rate in numerical results equals $\sum \alpha_i^2$.

In summary, we have confirmed that the results from the theoretical analysis agree with those obtained through numerical calculation. This supports the conclusion that our theoretical analysis is applicable to the actual Bitcoin network.

\begin{table}[t]
\caption{Comparison of theoretical and numerical slopes.}
\label{comparisonforksloop}
\begin{center}
\begin{tabular}{|c|c|c|}
\hline
$d/T$ & Theoretical result & Numerical result \\ \hline
0.01  & 0.0199003          & 0.019896         \\ \hline
0.04  & 0.0784211          & 0.0783533        \\ \hline
0.07  & 0.135212           & 0.135011         \\ \hline
\end{tabular}
\end{center}
\end{table}

\subsection{Insights} \label{insight}
Equation~\ref{mprapprox} offers several important insights. In this section, we elaborate on each of them in detail.

First, regarding whether the TRGR, the central focus of this study, holds in Bitcoin: we find that it does hold in the sense that a higher hashrate proportion leads to a higher mining profit rate. In particular, the mining profit rate depends linearly on the hashrate proportion.

\begin{figure}[tb]
  \centering    
  \includegraphics[width=\linewidth]{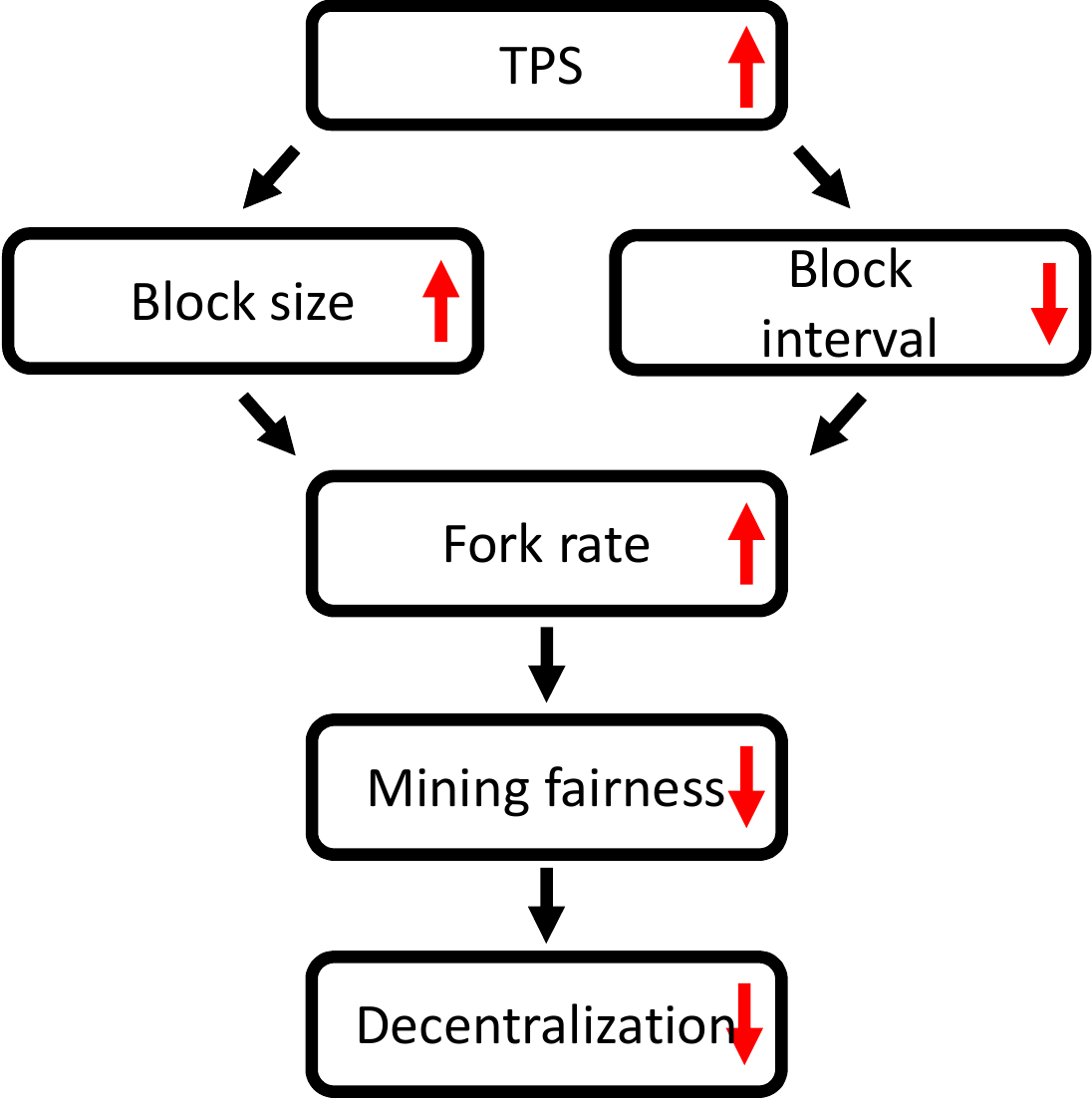}
  \caption{The trade-off between transaction processing capacity and decentralization. Increasing the block size or shortening the block interval improves transaction throughput, but also leads to more frequent blockchain forks. While more forks do not necessarily imply a stronger TRGR and thus may not always reduce decentralization, Equation \ref{mprapprox} clearly shows that improving throughput inherently strengthens TRGR, and consequently undermines decentralization.}
  \label{dillemmatd}
\end{figure}

Second, Equation~\ref{mprapprox} clarifies the trade-off between Bitcoin’s transaction processing capacity and its degree of decentralization (Figure~\ref{dillemmatd}). Focusing on the slope of \ref{mprapprox}, we see that it is $2f$, where $f$ is defined as $1 - \exp(-d/T)$. In other words, $f$ increases with greater block propagation delay $d$ or shorter block generation interval $T$. Since larger block sizes increase propagation delay, $f$ tends to increase as the system’s transaction processing capacity increases. An increase in $f$ signifies a stronger TRGR tendency, thus undermining decentralization.

Next, we consider the impact of hashrate distribution. Equation \ref{mprapprox} shows that the zero-crossing point of the mining profit rate is equal to the sum of the squares of the hashrate proportions. This value increases when the hashrate distribution becomes more skewed, indicating that the mining efficiency of the system decreases as the distribution becomes more centralized.

We then consider the effect of the tie-breaking rule. At first glance, Equation \ref{mprapprox} seems to suggest that the tie-breaking rule has no effect on the mining profit rate. However, Sakurai et al.'s block reward formula \ref{blockreward} suggests that the last-generated rule contributes the most to improving fairness, while the first-seen rule degrades it the most. This discrepancy arises because, in deriving \ref{mprapprox}, a term with coefficient $f^2$—which contains the influence of the tie-breaking rule—was approximated away. Thus, the effect of the tie-breaking rule is embedded in the term that was neglected. Conversely, this also implies that in systems with larger $f$, the mining profit rate is more sensitive to the choice of tie-breaking rule.

\section{Variable Block Propagation Delays Between Different Miners} \label{variable analysis}

The previous analyses were conducted under the assumption that block propagation delays between different miners are fixed. In this section, we investigate the robustness of those results under conditions where this assumption does not necessarily hold. Specifically, we first compare the cases where block propagation delays are fixed and where they are randomized. Then, from the perspective of economic rationality, we examine whether TRGR is preserved when miners are allowed to strategically manipulate block propagation delays.

\subsection{Randomizing Block Propagation Delays}

We compare the cases with randomized and fixed block propagation delays to examine how the structure of TRGR changes under randomness.

\subsubsection{Model of the Block Propagation Protocol} \label{gossipmodel}

In Bitcoin, blocks are propagated via a gossip-based flooding protocol, meaning that each miner forwards blocks to its neighbors. Under this protocol, the dissemination of a block throughout the system is commonly modeled using the following logistic differential equation:
\begin{align}
  \frac{dI}{dt} = \beta I(N - I),
\end{align}
where $\beta$ is a constant, $I(t)$ denotes the number of miners that have received the block by time $t$, and $N$ is the total number of miners.

Under the initial condition $I(0) = 1$, the solution of this equation is given by:
\begin{align}
  I(t) = \frac{N}{1 + (N-1) \exp^{-\beta N t}}.
\end{align}

Considering $I(t)/N$ as the cumulative distribution function (CDF) of a probability distribution, this corresponds to a logistic distribution. Therefore, the expected value of the block propagation delay is given by $\ln(N-1)/(\beta N)$.

\subsubsection{Settings} \label{randomsetting}

We set the number of miners to $N = 1000$ and the average block generation interval to $T = 600$ seconds. The distribution of hashrates among miners was based on the actual distribution observed in Bitcoin (see Figure~\ref{bitcoindis}).

Block propagation delays between miners were randomly drawn from a logistic distribution with a mean of 6.

\subsubsection{Results}

\begin{figure}[tb]
  \centering    
  \includegraphics[width=\linewidth]{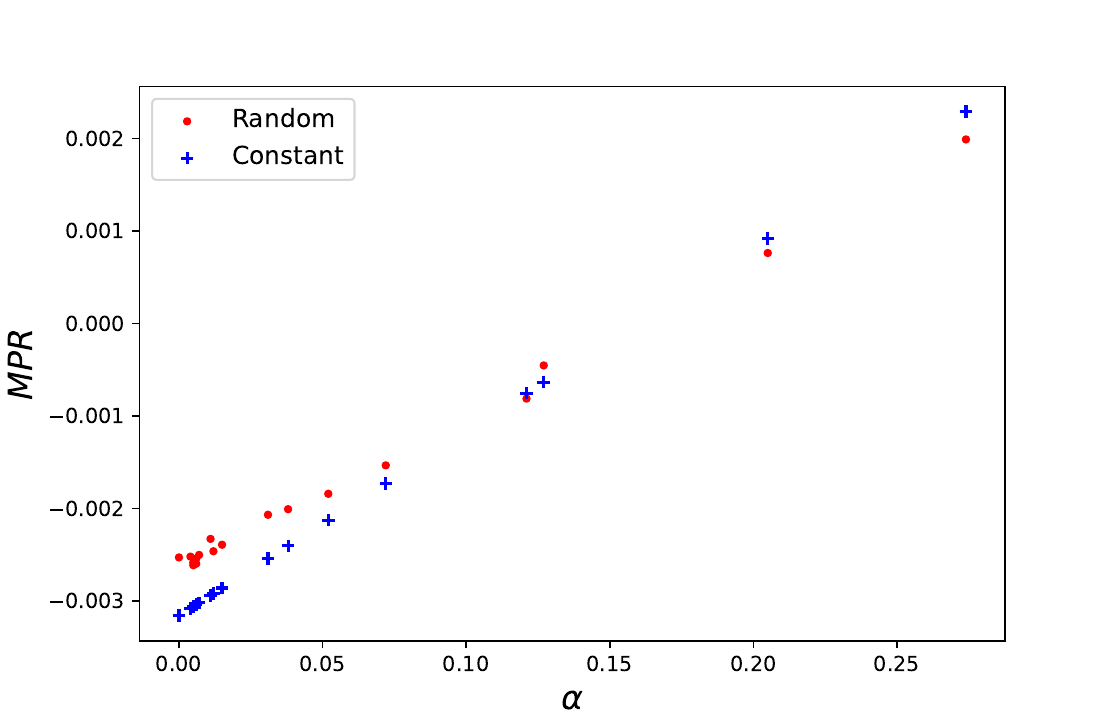}
  \caption{Comparison of the average mining profit rate between the case where block propagation delays between miners are randomized and the case where they are fixed.}
  \label{expectedmpr}
\end{figure}

We conducted 100 simulations of mining profit rates under randomized block propagation delays between miners. Figure~\ref{expectedmpr} shows the comparison of the results with those from the fixed-delay setting. The figure suggests that the average values are approximately the same in both cases. This is likely because $F_{ij}$, which depends on $T_{ij}/T$, can be approximated linearly with respect to $T_{ij}$ when $T_{ij}/T$ is sufficiently small.

\begin{figure}[tb]
  \centering    
  \includegraphics[width=\linewidth]{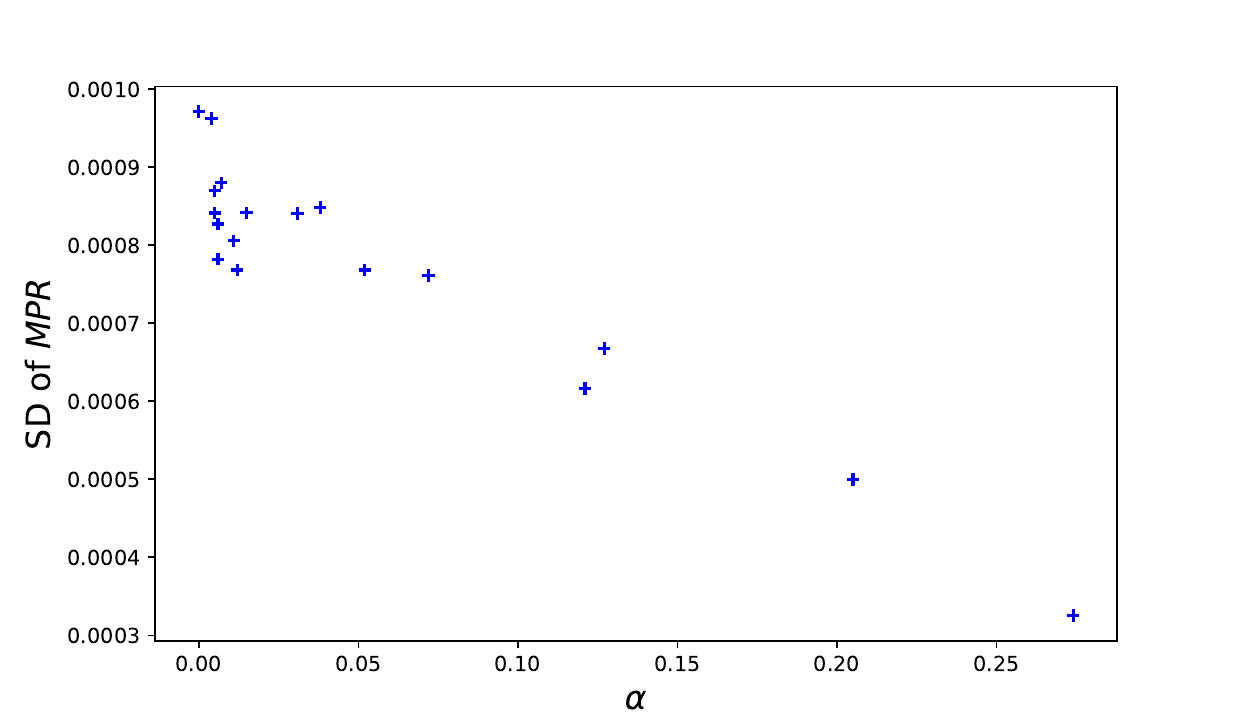}
  \caption{Standard deviation of mining profit rates under randomized block propagation delays between miners.}
  \label{sdmpr}
\end{figure}

Figure~\ref{sdmpr} shows the standard deviation of each miner's mining profit rate under the randomized setting. As seen in the figure, miners with smaller proportions of hashrate tend to exhibit higher standard deviation in their mining profit rates. This indicates that smaller miners are more susceptible to fluctuations in block propagation delays.

\subsection{From the Perspective of Economic Rationality}

We investigate whether TRGR is preserved under variable block propagation delays from the perspective of economic rationality. More specifically, we analyze the case under the assumption that each miner is allowed to manipulate block propagation delays as part of their strategy to maximize their own mining profit rate.

\begin{figure*}[tb]
  \begin{minipage}[tb]{0.33\linewidth}
    \centering
    \includegraphics[scale=0.35]{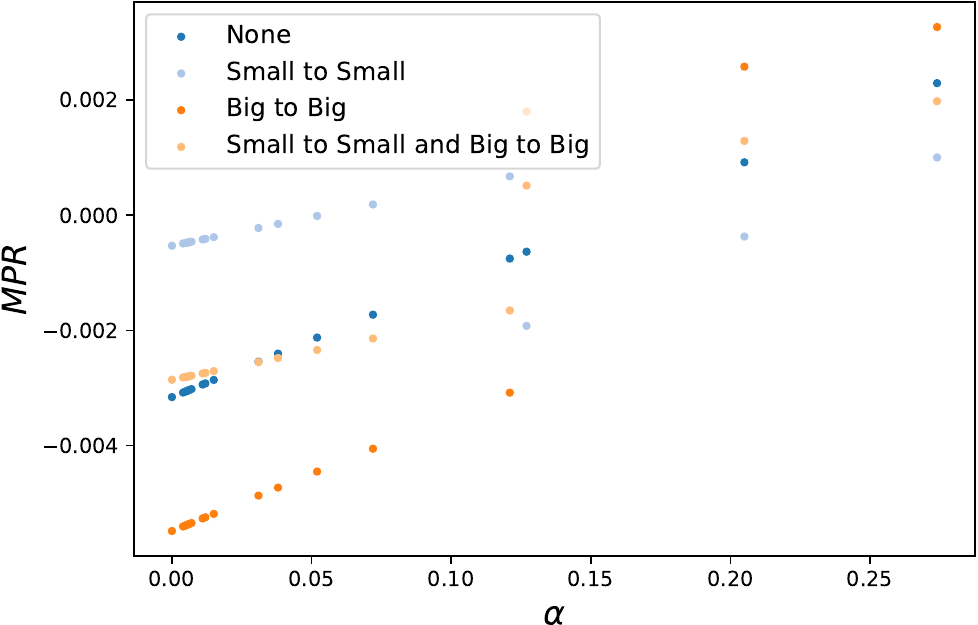}
    \subcaption{First-seen rule.}
  \end{minipage}
  \begin{minipage}[tb]{0.33\linewidth}
    \centering
    \includegraphics[scale=0.35]{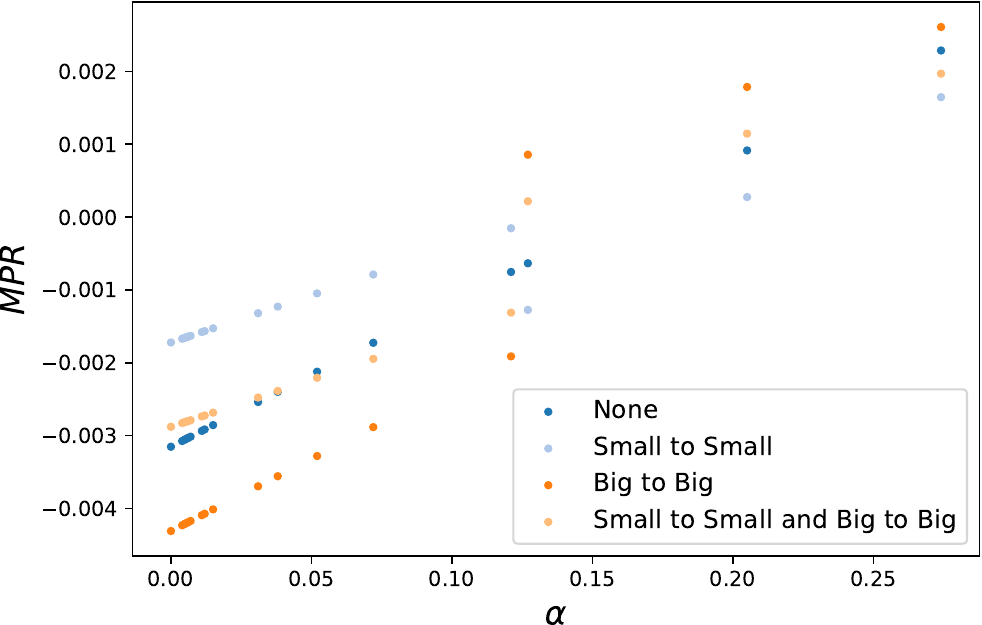}
    \subcaption{Random rule.}
  \end{minipage}
  \begin{minipage}[tb]{0.33\linewidth}
    \centering
    \includegraphics[scale=0.35]{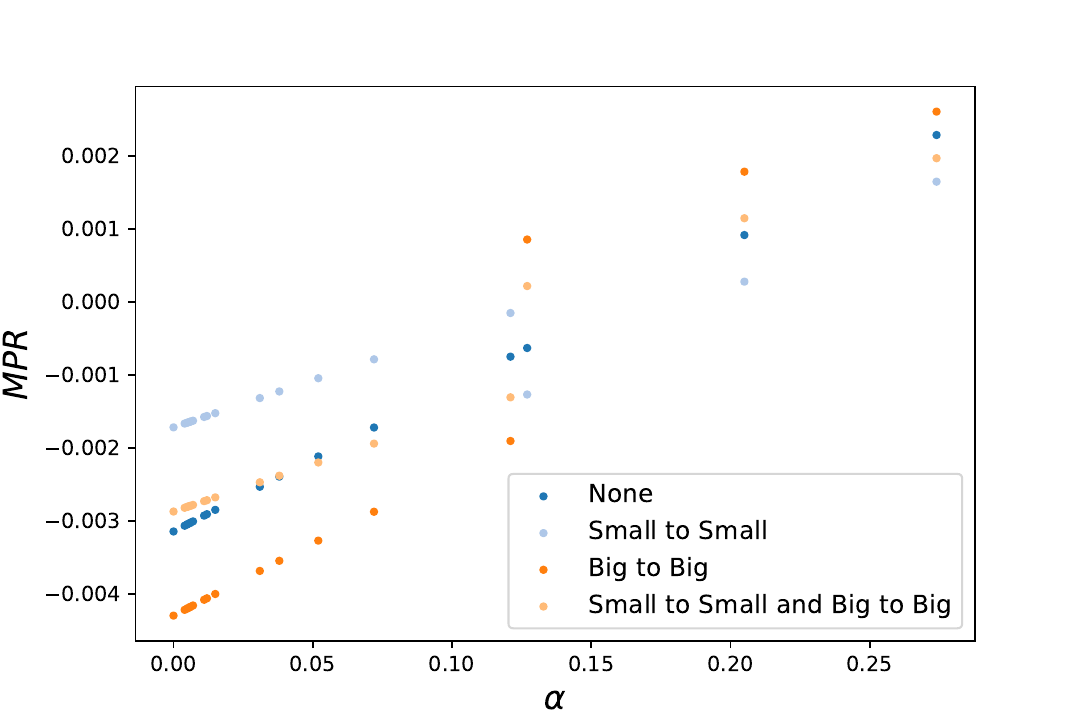}
    \subcaption{Last-generated rule.}
  \end{minipage}
  \caption{Numerical results of the strategic interactions.}
  \label{gameresult}
\end{figure*}

\subsubsection{Setup}

We considered $N = 1000$ miners and set the average block generation interval to $T = 600$ seconds. The hashrate distribution was based on the actual distribution observed in the Bitcoin network (see Figure~\ref{bitcoindis}). 

It is not easy to analyze the full strategy space of all miners. Therefore, we simplified the setting as follows. Each miner's strategy consists of choosing the propagation delay for blocks it sends. The block propagation delay between any pair of miners was set to either $d = 3$ (fast propagation) or $d = 6$ (slow propagation). We assumed that all miners were equally capable of accelerating or delaying propagation. In practice, selecting between fast and slow propagation is relatively easy. To slow propagation, a miner can delay transmission or continue mining as if it has not received the block. To accelerate propagation, a miner can establish a direct connection or use high-bandwidth modes such as Compact Block Relay~\cite{bip152}.

Miners were divided into two groups: large and small miners. Specifically, the group of miners accounting for the top 50\% of the total hashrate was defined as large miners. All miners within each group were treated homogeneously. For example, if one large miner chooses to propagate blocks quickly to another large miner, then all large miners are assumed to do the same toward all other large miners.

The utility of each group was defined as the aggregated mining profit rate of the miners in that group.

\subsubsection{Results}

First, we observe that fast block propagation from large miners to small miners, or vice versa, is not a rational outcome. This is because accelerating block propagation requires agreement from both the sender and the receiver, and Bitcoin mining is a zero-sum game. If a large miner propagates a block faster to a small miner and gains a profit, the small miner must lose an equivalent amount. In this case, small miners have no incentive to cooperate in accelerating propagation, and mutual agreement cannot be established.

From this reasoning, the only viable strategies for large miners are whether to propagate blocks quickly to other large miners. Similarly, small miners can only decide whether to accelerate propagation to other small miners.

Figure~\ref{gameresult} shows the outcomes of each group's strategies. The results indicate that accelerating block propagation within one’s own group improves that group's mining profit rate. Furthermore, a Nash equilibrium is reached when both groups choose to accelerate block propagation within their own group. In this equilibrium, the TRGR structure remains intact, indicating that the emergence of TRGR is consistent with economic rationality.

\section{Discussion}

\subsection{Mitigating TRGR}

In this section, we discuss potential approaches to mitigating TRGR.

One straightforward approach is to reduce block propagation delays. As suggested by Eq.~\ref{mprapprox}, reducing block propagation delays leads to a smaller slope, thereby improving the overall mining fairness of the system. A key advantage of this approach is its compatibility with the existing system. Improvements to block propagation protocols can be implemented without requiring changes to the Bitcoin protocol itself~\cite{bip-0125, Graphene}. Such compatibility is especially important for Bitcoin, which is a currency system and favors conservative system operations. On the other hand, implementing these propagation protocols is not a fully local operation; it requires coordination between communication peers.

Another promising approach is to modify the tie-breaking rule. Among the possible options, the first-seen rule leads to the most unfair outcomes. Because Bitcoin is a decentralized system, miners with higher hashrates tend to have a higher proportion of round initiation. Under the first-seen rule, miners who initiate rounds more frequently gain an advantage in tie situations, amplifying the inequality. In contrast, the random rule and the last-generated rule help suppress disparities in proportions of round initiation and thereby improve mining fairness.

Compared to the propagation delay improvement approach, modifying the tie-breaking rule has the benefit of better local compatibility. For instance, the method proposed by Sakurai et al.~\cite{sakurai2024fullylocallastgeneratedrule} can be implemented and operated entirely locally by each miner. However, as indicated by Eq.~\ref{mprapprox}, the impact of modifying the tie-breaking rule on mining fairness is generally smaller than that of improving block propagation delays. 

A third approach is to reward stale blocks—i.e., blocks that are not included in the main chain. In principle, since each miner's number of generated blocks aligns with its hashrate proportion, this mechanism can achieve mining fairness. Similar approaches have been proposed in prior studies to counter attacks that degrade mining fairness~\cite{majorityisnotenough, FruitChains, Prism}. However, this approach has two major drawbacks. The first is compatibility with Bitcoin. As a currency system, Bitcoin tends to favor conservative upgrades. Most existing proposals for rewarding stale blocks break backward compatibility, making them difficult to deploy in practice. There are more compatible alternatives, such as decentralized mining pools~\cite{p2pool, SmartPool, sakurai2025fiberpoolleveragingmultipleblockchains}, but research into incentive mechanisms for participating in such pools is still limited, and it remains unclear how widely these systems can be adopted.

The second issue is that rewarding stale blocks may increase the incentive to attack the Bitcoin network. If forked blocks receive rewards, then blocks intentionally generated for attacks would also be rewarded. Indeed, prior research~\cite{UncleBlockMechanismEffectonEthereumSelfishndStubbornMining} has shown that Ethereum~\cite{ethereum}, which partially adopts this approach, faces an increased risk of selfish mining as a result.

\subsection{The Importance of Considering Proportions of Round Initiation in Mining Fairness Analysis}

Simulation experiments have shown that neglecting the effect of forks on proportions of round initiation can lead to mining profit estimation errors of nearly 100\%~\cite{sakurai2025modelbasedanalysisminingfairness}. The high accuracy of Sakurai et al.'s method stems mainly from the fact that their blockchain model is round-based, allowing forks to be treated formally. However, it was previously unclear why ignoring the influence of forks on proportions of round initiation causes such large errors.

To clarify this, we compute the mining profit rate under the assumption that proportions of round initiation are equal to hashrate proportions, i.e., ignoring the impact of forks. By proceeding with calculations as in Section~\ref{theoretical analysis}, we obtain:
\begin{align}
MPR_i = f \left(\alpha_i - \sum_{j \in V} \alpha_j^2\right).
\end{align}
In contrast, when considering the impact of forks on proportions of round initiation, the profit rate becomes:
\begin{align}
MPR_i = 2f \left(\alpha_i - \sum_{j \in V} \alpha_j^2\right).
\end{align}
This result shows that the effect of forks on proportions of round initiation is as significant as their other effects on mining fairness.

\subsection{Limitations}

The simplified form of TRGR, as expressed in Eq.~\ref{mprapprox}, is derived under the assumption of fixed block propagation delays. While we have shown that TRGR still holds under economic rationality even when delays are variable, the setting used in this analysis is deliberately limited to ensure tractability. For example, we grouped miners into two categories—those accounting for the top 50\% of total hashrate and others—and treated all miners within each group identically. Investigating how far such assumptions can be relaxed while preserving TRGR is an important direction for future work.

In addition, our study neglects the distributed nature of mining pools. Although mining pools behave like single miners on the Bitcoin network, they are in fact composed of multiple cooperating miners who are geographically distributed. Due to this distributed nature, discarded blocks that are not recorded on the blockchain may still be generated, which could affect mining fairness.

However, we argue that the effect of intra-pool distribution can be ignored. During synchronization within a pool, each miner only receives minimal information necessary for mining from the pool server. This data is significantly smaller than what is needed to participate in the Bitcoin network itself, so synchronization within the pool is fast. Therefore, the impact of intra-pool delays is expected to be much smaller than inter-miner delays on the public network.




\section{Conclusion}
In this study, we theoretically demonstrate TRGR in Bitcoin under the assumption of fixed block propagation delays between miners, using a significantly more precise method than previous research. We also analyze the impact of block propagation delay, hashrate, and the tie-breaking rule on TRGR. Furthermore, we validate TRGR in a setting where block propagation delays are not fixed.

\bibliography{hoge} 
\bibliographystyle{IEEEtran.bst} 

\end{document}